\documentclass[aps,prl,showpacs,twocolumn]{revtex4}
\usepackage{dcolumn}
\usepackage{graphicx}

\begin{document}
\raggedbottom

\title{Optical pumping of trapped neutral molecules by blackbody radiation}
\author{Steven Hoekstra}
\author{Joop J.~Gilijamse}
\author{Boris Sartakov}
\altaffiliation[Permanent address: ]{General Physics Institute,
Russian Academy of Sciences, Vavilov street 38, Moscow 119991,
Russia.}
\author{Nicolas Vanhaecke}
\altaffiliation[Present address: ]{Laboratoire Aim\'{e} Cotton,
CNRS II, Campus d'Orsay, Bˆatiment 505, 91405 Orsay cedex,
France.}
\author{Ludwig Scharfenberg}
\author{Sebastiaan Y.~T.~van de Meerakker}
\author{Gerard Meijer}
\affiliation{Fritz-Haber-Institut der Max-Planck-Gesellschaft,
Faradayweg 4-6, 14195 Berlin, Germany}

\begin{abstract}
Optical pumping by blackbody radiation is a feature shared by all
polar molecules and fundamentally limits the time that these
molecules can be kept in a single quantum state in a trap. To
demonstrate and quantify this, we have monitored the optical
pumping of electrostatically trapped OH and OD radicals by
room-temperature blackbody radiation. Transfer of these molecules
to rotationally excited states by blackbody radiation at 295 K
limits the $1/e$ trapping time for OH and OD in the
$X^{2}\Pi_{3/2},v''=0,J''=3/2(f)$ state to 2.8 s and 7.1 s,
respectively.
\end{abstract}

\pacs{33.55.Be, 33.80.Ps, 44.40.+a}
\maketitle

In his 1917 paper Einstein showed~\cite{EINSTEIN1917B} that even
in the absence of collisions the velocity distribution of a
molecular gas takes on a Maxwellian distribution due to the
momentum transfer that takes place in the absorption and emission
of blackbody radiation. The absorbed and emitted photons optically
pump the rotational and vibrational transitions, resulting in
thermal distributions over the available states. The rotational
temperature of the CN molecule in interstellar
space~\cite{MCKELLAR1941}, for example, is the result of optical
pumping by the cosmic microwave
background-radiation~\cite{PENZIAS1965A}.

The influence of blackbody radiation on atoms and molecules is in
general small and it is rare that it can be observed directly in
laboratory experiments. However, in a number of cases the
interaction with blackbody radiation is experimentally observable
and important. The first dynamical effects of blackbody radiation
on the population of atomic levels were noticed when studying the
lifetime of highly excited Rydberg states in
atoms~\cite{GALLAGHER1979}. Atoms in these states can have dipole
moments of thousands of Debye, and have sufficient spectral
overlap with the spectrum of room-temperature blackbody radiation.
The excitation (and ionization) rates can be on the order of 1000
s$^{-1}$, implying that the effect can already be observed on a
$\mu$s timescale.

The excitation rates in ground state atoms and molecules are
generally much lower, and therefore require a longer interaction
time to be observed. Only with the development of ion traps,
together with a sufficient reduction of collisional energy
exchange (i.e. a good vacuum at room temperature), could the
photodissociation of molecular ions and clusters by blackbody
radiation be directly observed~\cite{DUNBAR1991,THOLMANN1994}.
Ions in storage rings are also trapped long enough for the
interaction with blackbody radiation to be
noticeable~\cite{HECHTFISCHER1998}.

The effect of blackbody radiation on neutral molecules in a trap
has until now been left experimentally unexplored, partly because
the conditions to observe the effect were not met, and partly
because the importance of this effect was not always realized.
Polar molecules generally have strong vibrational and/or
rotational transitions in the infrared region of the spectrum. As
a result they can relatively easily be optically pumped by
room-temperature blackbody radiation, and this fundamentally
limits the time that these molecules can be kept in a single
quantum state in room temperature traps. This has important
implications for cooling schemes which aim at increasing the
phase-space density of trapped neutral polar molecules.

In the rapidly growing research field of cold molecules, these
polar molecules are of special interest. When quantum degeneracy
is reached with polar molecules it will be much different from
atomic Bose-Einstein condensates or degenerate Fermi gasses due to
the dipole-dipole interaction~\cite{BARANOV2002O, STUHLER2005}.
Trapped polar molecules also hold promise for use in quantum
information systems~\cite{ANDRE2006}. Furthermore a search for a
permanent electric dipole moment of the electron, which would
result in the violation of time reversal symmetry, is being done
using polar molecules~\cite{HUDSON2002} because an external
electric field applied to a polar molecule leads to a hugely
magnified internal field. Quantum chemistry and collisions at low
energy~\cite{GILIJAMSE2006} are other examples of the
possibilities with cold polar molecules.

Many of these exciting experiments require a dense and cold sample
of trapped polar molecules. Therefore effort in the research on
cold molecules is currently aimed at increasing the phase-space
density of the trapped molecules~\cite{DOYLE2004}. A natural
approach to try to reach this goal is to adapt the cooling
techniques such as evaporative cooling~\cite{KETTERLE1996} and
sympathetic cooling~\cite{MODUGNO2001}, but these cooling
techniques require usually trapping times of 10 s or more.
Therefore an investigation of blackbody radiation as a limiting
factor for the trapping time of polar molecules is of importance.
Also for future studies of collisions between trapped molecules a
quantitative understanding of all trap loss mechanisms is
essential. In this Letter we experimentally quantify trap loss due
to blackbody radiation for neutral polar molecules stored in a
room-temperature trap, using the OH radical as a model system.
Cold packets of OH radicals are produced using the Stark
deceleration and trapping technique\cite{BETHLEM1999, HEINER2006}.

A natural approach to experimentally study the effect of blackbody
radiation on the trapped molecules would be to vary the
temperature of the trapping apparatus, thereby changing both the
spectral distribution and the intensity of the blackbody
radiation. A disadvantage of this is that normally a change of the
temperature will also influence the vacuum conditions, changing
the collision rate with the residual gas. The alternative approach
we have taken is to compare the room-temperature trapping of two
isotopic variants of the same molecule. Under the assumption that
these isotopomers have the same collisional properties, the trap
losses due to blackbody radiation and collisions with background
gas can be disentangled.

\begin{figure}
\resizebox{\linewidth}{!}{\includegraphics{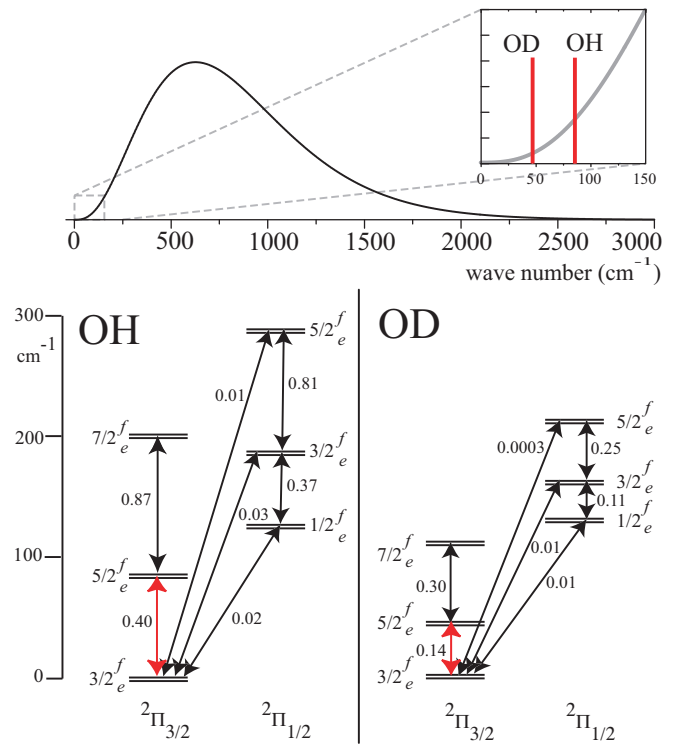}}
\caption{\label{fig:levels} In the upper part, the
room-temperature spectrum of blackbody radiation is shown. The
relevant region for rotational excitation of OH and OD radicals is
shown, with sticks indicating the lowest rotational excitation
frequency. In the lower part the relevant rotational levels of OH
and OD are depicted. Each rotational level is split into two
$\Lambda$-doublet components with opposite parity, denoted by $f$
and $e$. The size of the splitting is exaggerated for clarity. The
most relevant transition rates (in s$^{-1}$) due to blackbody
radiation connecting the $f$-components of rotational levels are
indicated next to the arrows.}
\end{figure}

The molecules we have compared are the OH and OD radical. There is
sufficient intensity in the low-energy side of the blackbody
radiation spectrum that these molecules can be rotationally
excited; the vibrational energy splitting is so large that
vibrational excitation is negligible. The energies of the relevant
rotational states of OH and OD in the $X^{2}\Pi$-state are
depicted in figure~\ref{fig:levels}. Each rotational level is
split into two $\Lambda$-doublet components with opposite parity,
denoted by $f$ and $e$. In an electric field these components are
mixed. In the electric field, each $J$-state is split into
$(2J+1)$ $M_{J}\Omega$ components. The $f$ component in zero-field
corresponds with $(2J+1)/2$ low-field seeking components in an
electric field; the $e$ component corresponds with $(2J+1)/2$
high-field seeking components.

Blackbody radiation can excite transitions with $\Delta M_{J}=\pm
1,0$. Molecules trapped in low-field seeking states can be pumped
to high-field seeking states, leading to trap loss. Because the
rotational constant $B$ is almost a factor of two larger for OH
than for OD, the intensity of the room-temperature blackbody
radiation is much higher at the excitation energy of the first
rotational excitation for OH molecules than it is for OD
molecules. The most important transition rates due to the
blackbody radiation connecting the $f$-components of the lowest
rotational levels are indicated next to the arrows in
figure~\ref{fig:levels}. The difference in the excitation rates
leads us to expect significantly shorter trapping times for OH
than for OD radicals.

\begin{figure}
\resizebox{\linewidth}{!}{\includegraphics{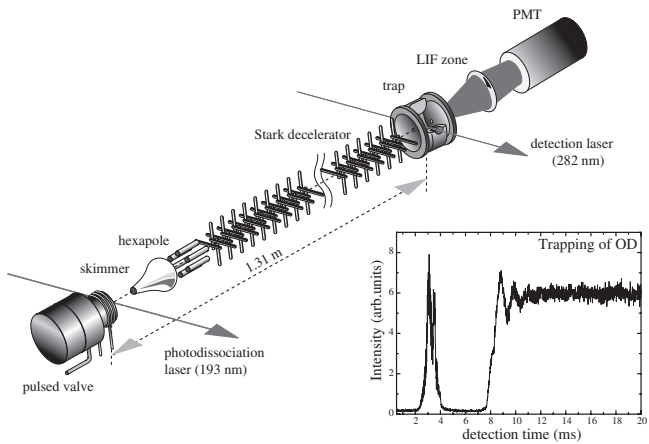}}
\caption{\label{fig:setup}Scheme of the experimental setup. OH
(OD) radicals are created by photodissociation of supersonically
expanded (deuterated) nitric acid, and subsequently decelerated in
a Stark decelerator. The slowed molecules are trapped in an
electrostatic trap, and the population in the different rotational
states is probed by laser-induced fluorescence detection. In the
inset the arrival of undecelerated OD molecules followed by the
steady signal of the trap-loaded OD molecules is shown, up to 20
ms after production of the molecules. We have detected the trapped
molecules up to 15 s.}
\end{figure}

The production, deceleration and trapping of OH has already been
described elsewhere~\cite{VANDEMEERAKKER2005}; here we summarize
the main procedure. The experimental setup is schematically
depicted in figure~\ref{fig:setup}. The OH (OD) radicals are
created, in separate experiments, by photodissociation of
HNO$_{3}$ (DNO$_{3}$), co-expanded with xenon as a carrier gas in
a pulsed supersonic expansion. After the expansion both OH and OD
are mostly in the electronic, vibrational and rotational ground
state $X^{2}\Pi_{3/2}$ (v=0, J=3/2). The $M_{J} \Omega=-9/4$
component of the $3/2(f)$ state is the strongest low-field-seeking
state and is focussed by a hexapole lens into the decelerator.
Using 108 deceleration stages a part of the beam is brought to a
standstill, and subsequently trapped in an electrostatic
quadrupole trap. In these experiments the OD radical, which is
predicted to have interesting collisional
properties~\cite{AVDEENKOV2005}, has been trapped for the first
time. Due to their similar mass and dipole moment OH and OD are
equally well decelerated and trapped. We estimate the trapped
number density to be about $10^{7}$ molecules/cm$^{3}$, and have
not found any indication of collisions between the trapped
molecules. The trapped molecules, probed by
laser-induced-fluorescence directly after switching off the
trapping voltages, are initially exclusively in the $M_{J}
\Omega=-9/4$ component of the $3/2(f)$ state.

\begin{figure}
\resizebox{\linewidth}{!}{\includegraphics{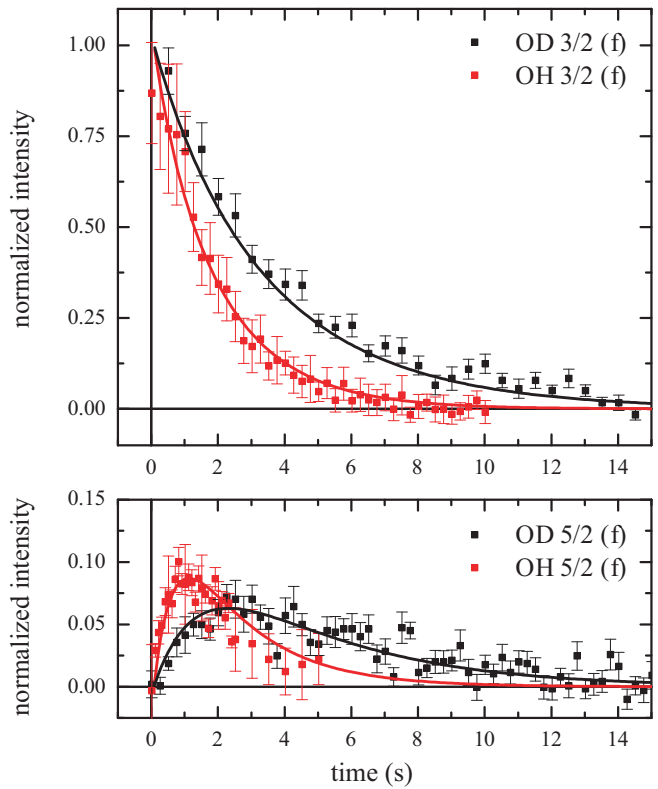}}
\caption{\label{fig:trappingtime}The population of
electrostatically trapped OH and OD radicals in the $3/2(f)$ and
$5/2(f)$ levels as a function of time. The squares are the
measured data - red (grey) for OH, black for OD. The curves are
the outcome of the rate equation model using a constant background
gas collision rate of 0.17 s$^{-1}$.}
\end{figure}

We have measured, for OH and OD, the population trapped in the
$3/2(f)$ state, up to a trapping time of 15 seconds. The results
are shown in the upper part of figure~\ref{fig:trappingtime}. The
OD molecules remain, as expected, approximately a factor of two
longer in the $3/2(f)$ state than the OH molecules. To interpret
the data we have used a rate equation model, describing the time
evolution of the population in the rotational states. The curves
in figure~\ref{fig:trappingtime} show the outcome of this model.
Molecules can be pumped from the $3/2(f)$ state to the $5/2(f)$
state, in which they can remain trapped. The time-evolution of the
$5/2(f)$ population follows from the rate-equation model. The
results of the measurement of the population in this state are
shown in the lower part of figure~\ref{fig:trappingtime}. About 5
\% of the original OD population accumulates in the $5/2(f)$ state
after 3 seconds of trapping. For OH the maximum in the $5/2(f)$
population occurs after 1 second of trapping. From the $5/2(f)$
state the molecules can be pumped to other rotational states
again, spontaneously decay back to the $3/2(f)$ state or be lost
due to collisions with the background gas.

For the model, the transition strengths of all allowed transitions
between the $M_{J}$-components of the rotational states (for both
$\Omega$ manifolds, up to $J=7/2$) in OH and OD are calculated.
The energy differences caused by hyperfine effects, the Stark
shift of the levels in the electric field of the trap and the
$\Lambda$-doubling itself are so small compared to the rotational
spacing that these are neglected in the calculation of the
transition rates. Using the calculated transition strengths and
the spectrum of the blackbody radiation at 295 K the transition
rates due to the blackbody radiation are obtained, as shown in
figure~\ref{fig:levels}. Because not all $M_{J}$-components have
an equal Stark shift the different trapping depths for these
states were taken into account. Molecules excited to the $J \geq
7/2$ states are considered lost from the trap. Stimulated emission
was also included. Loss due to collisions with residual background
gas, assumed to be independent of the rotational level, is taken
into account by a constant loss rate. We find best agreement
between the model and the data using a background-gas loss rate of
$0.17$ s$^{-1}$\footnote{A better fit can be obtained by using a
time-dependent background gas loss rate $r(1+ae^{-t/\tau})$, with
$r=0.13$ s$^{-1}$, $a=1.0$ and $\tau=1.0$ s, to account for the
pressure decrease following the injection of gas at the beginning
of the experimental cycle.}. To verify whether this loss rate can
be explained by collisions with a xenon background gas we have
used the recently calculated cross-section for Xe-OH
collisions~\cite{GILIJAMSE2006}, taken to be identical for OH and
OD. For a room temperature gas of xenon atoms colliding with
trapped OH molecules the average collision energy is 30 cm$^{-1}$,
leading to an elastic cross section of $500 \pm 50$ $\AA^{2}$;
rotational excitation of trapped OH and OD by the xenon atoms can
safely be neglected. Because the trap depth is only about 1
cm$^{-1}$ almost every collision will lead to trap loss. Using a
partial xenon pressure of $5 \cdot 10^{-9}$ mbar we find a
collision rate of 0.14 s$^{-1}$, which is consistent with the rate
found from the comparison of the model with the data.

\begin{table}
\caption{\label{tab:list}Pumping rates due to blackbody radiation
at two different temperatures, out of the specified initial state,
for a number of polar molecules.}
\begin{ruledtabular}
\begin{tabular}{l|l|l|l}
System & Initial state &\multicolumn{2}{l}{Pumping rate (s$^{-1}$)} \\
 & & 295 K & 77 K \\
\hline
OH/OD & {$X^{2}\Pi_{3/2},J=\frac{3}{2}$} & 0.49/0.16 & 0.058/0.027\\
NH/ND & {$a^{1}\Delta,J=2$} & 0.36/0.12 & 0.042/0.021\\
NH/ND & {$X^{3}\Sigma^{-},N=0,J=1$} & 0.12/0.036 & 0.025/0.0083\\
NH$_{3}$/ND$_{3}$ & {$\tilde{X}^{1}A_{1}^{'},J=1,|K|=1$} & 0.23/0.14 & 0.019/0.0063\\
SO & {$X^{3}\Sigma^{-},N=0,J=1$} & 0.01 & $<10^{-3}$\\
$^{6}$LiH/$^{6}$LiD & $X^{1}\Sigma^{+},J=1$ & 1.64/0.81 & 0.31/0.11\\
CaH/CaD & {$X^{2}\Sigma^{+},N=0,J=\frac{1}{2}$} & 0.048/0.063 & 0.0032/$<10^{-3}$\\
RbCs & {$X^{1}\Sigma^{+},J=0$} & $<10^{-3}$ & $<10^{-3}$\\
KRb & {$X^{1}\Sigma^{+},J=0$} & $<10^{-3}$& $<10^{-3}$\\
CO & {$a^{3}\Pi_{1,2},J=1,2$} & 0.014/0.014 & $<10^{-3}/<10^{-3}$\\
\end{tabular}
\end{ruledtabular}
\end{table}

We have calculated the blackbody pumping rates for a number of
other polar molecules as well. These molecules are well suited for
trapping, using the various currently available
techniques~\cite{DOYLE2004}. For the selected molecules the
blackbody pumping rate out of the specified initial quantum state
is compared, for two different temperatures, in
table~\ref{tab:list}. The initial state is in most cases the
electronic, vibrational and rotational ground state. The rate for
OH is slightly larger than the sum of the rates indicated in
figure~\ref{fig:levels} due to the effect of hyperfine structure
that was included in these calculations.

The room-temperature pumping rate for many of the listed molecules
is comparable to that of OH and OD. LiH, with its favorable
Stark-effect to mass ratio and the existence of 4 isotopomers
often considered an ideal candidate molecule~\cite{TOKUNAGA2006},
is especially sensitive to blackbody radiation: even at 77 K it
can only be trapped for a few seconds. At 4 K the rates for all
molecules are smaller than $10^{-3}$. Depending on the molecule
the pumping rate is either dominated by rotational transitions
(OH/OD,NH/ND,LiH/LiD), by vibrational transitions (CO,RbCs,KRb) or
by a combination of both. CO, RbCs and KRb have very small
blackbody pumping rates.

In this Letter we have experimentally studied trap loss of neutral
polar molecules due to room-temperature blackbody radiation, using
the OH radical as a model system. By comparing the trapping times
of OH and OD radicals in an electrostatic quadrupole trap, the
individual contribution of blackbody radiation and collisions with
background gas could be quantified. Loss due to blackbody
radiation is a major limitation for the room-temperature trapping
of OH and OD radicals. If the vacuum conditions were improved such
that losses due to collisions with background gas were completely
removed, the $1/e$ trapping time would increase to 7.1 s for OD
and 2.8 s for OH. The trapped molecules will have to be shielded
from thermal radiation if longer trapping times are required. We
have shown that these limitations are shared by a large class of
polar molecules, which has profound consequences for the
implementation of further cooling schemes.

\bibliography{references}
\end{document}